\documentclass[10pt,aps,prd,nofootinbib,superscriptaddress,twocolumn]{revtex4}
\pdfoutput=1
\usepackage{amssymb,amsmath,latexsym,graphics, graphicx,epsfig,multirow,comment,hyperref,appendix} 

\newcommand{\gsim}{\lower.7ex\hbox{$\;\stackrel{\textstyle>}{\sim}\;$}}
\newcommand{\lsim}{\lower.7ex\hbox{$\;\stackrel{\textstyle<}{\sim}\;$}}

\def\OO{{\cal O}}

\newcommand{\GeV}{\,\mathrm{GeV}}

\newcommand{\bef}{\begin{figure}[htbp]\begin{center}}
\newcommand{\eef}{\end{center}\end{figure}}

\begin{document}
\pagestyle{plain}

\title{
\begin{flushright}
\mbox{\normalsize SLAC-PUB-14369}\\
\mbox{\normalsize SU-ITP-11/03}
\end{flushright}
\vskip 15 pt

Jet Dipolarity: 
\\
Top Tagging with Color Flow}
\author{Anson Hook}
\affiliation{
Theory Group, SLAC, Menlo Park, CA 94025}
\affiliation{
Physics Department, Stanford University,
Stanford, CA 94305}
\author{Martin Jankowiak}
\affiliation{
Theory Group, SLAC, Menlo Park, CA 94025}
\affiliation{
Physics Department, Stanford University,
Stanford, CA 94305}
\author{Jay G. Wacker}
\affiliation{
Theory Group, SLAC, Menlo Park, CA 94025}

\begin{abstract}
A new jet observable, dipolarity, is introduced that can distinguish whether a pair of subjets arises from a color singlet source.  This observable
is incorporated into the {\tt HEPTopTagger} and is shown to improve discrimination between top jets and QCD jets for moderate to high $p_T$.   
\end{abstract}
 \maketitle

\section*{Introduction}

The impressive resolution of the ATLAS and CMS detectors means that a typical QCD jet at the LHC deposits energy in $\OO(10\!-\!100)$ calorimeter cells.  Such fine-grained calorimetry allows for jets to be studied in much greater detail than previously, with 
sophisticated versions of current techniques making it possible to measure more than just the bulk properties of jets ({\it e.g.\;}event jet multiplicities or jet masses).
One goal of the LHC is to employ these techniques to extend the amount of information available from each jet, allowing for a broader probe of the properties of QCD.
 The past several years have seen significant progress in developing such jet substructure techniques. A number of general purpose tools have been developed, including: 
(i) top-tagging algorithms  designed for use at both lower \cite{Plehn:2009rk,Plehn:2010st} and higher \cite{Kaplan:2008ie} $p_T$ as well as 
(ii) jet grooming techniques such as filtering \cite{Butterworth:2008sd}, pruning \cite{Ellis:2009me}, and  trimming \cite{Krohn:2009th}, which are designed to improve jet mass resolution.
Jet substructure techniques have also been studied in the context of specific particle searches, where they have been shown to substantially extend the reach of traditional search techniques in a wide variety of scenarios, including for example boosted Higgses \cite{Butterworth:2008sd,Hackstein:2010wk}, neutral spin-one resonances \cite{Katz:2010mr}, searches for supersymmetry \cite{Kribs:2010hp}, and 
many others \cite{Chen:2010wk,Falkowski:2010hi,Kribs:2009yh,Butterworth:2007ke,Englert:2010ud}.   
Despite these many successes, however, there is every reason to expect that there remains room for refinement of jet substructure techniques. 

Top tagging algorithms have reached a mature level of development in recent years
\cite{Plehn:2009rk,Plehn:2010st,Kaplan:2008ie,brooijmans,cmsnote3, vos,Thaler:2008ju,Rehermann:2010vq,Almeida:2008tp}.  
A variety of different algorithms employ
primarily kinematic observables like jet masses and the $W^\pm$ helicity angle in increasingly sophisticated
ways, allowing for efficient discrimination between top jets and ordinary QCD jets.  One direction that has received less attention is the use of 
observables that do not map onto the kinematics of hard partons.  This article introduces a new non-kinematic observable
that can be used in top tagging to gain additional background rejection and should have applications outside of top tagging.

A distinguishing aspect of hadronic top decays is that the jets from the $W^\pm$ decay belong to a color singlet configuration.  For 
$p_{T\, W} \!\gtrsim\! m_W$ these jets become close together and will often be clustered within a single jet.  The radiation pattern of the $W^\pm$ 
decay products, which is controlled by the color configuration, has a distinctive form with most of the radiation clustered in between the 
two jets.  This QCD analog of the Chudakov effect offers an additional handle for top discrimination on top of kinematic observables.  

\vspace{-.022in}
The organization of this article is as follows.  First the concept of color flow is briefly reviewed, and connection is made
to the jet observable pull \cite{Gallicchio:2010sw}.
Next the proposed ``dipolarity'' observable is introduced and explored in a general context.  Dipolarity is then incorporated into the 
{\tt HEPTopTagger}, and the performance of the modified tagger is tested on Monte Carlo event samples.  Finally, there is some discussion of 
the results as well as possible applications of dipolarity beyond top tagging. 

\begin{figure}[h] %  figure placement: here, top, bottom, or page
   \centering
  \includegraphics[width=3.5 in]{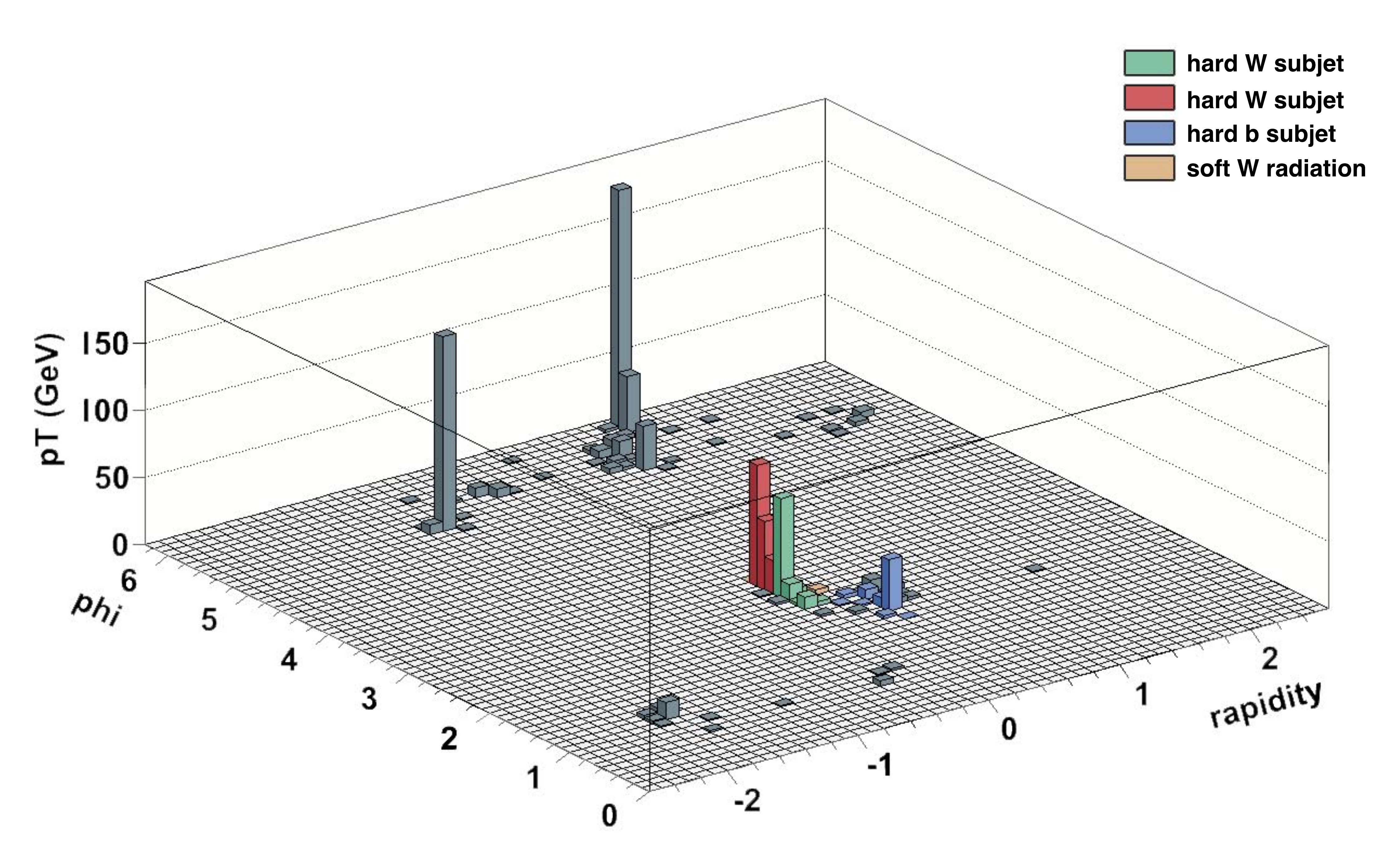}
   \caption{Legoplot for a top jet reconstructed by the {\tt HEPTopTagger}, with hard substructure identified
   by a combination of filtering and a fractional mass-drop criterion.  The orange cells correspond to soft radiation associated 
   with the $W^\pm$.}
   \label{fig:legoplot}
\end{figure}

\section*{Color flow and pull}

Within the context of top tagging, several jet observables have been defined that go beyond the kinematics
of hard partons.  These include a number of jet shape observables such as spherocity \cite{Thaler:2008ju},
planar flow \cite{Almeida:2008tp, Almeida:2008yp}, $N$-subjettiness \cite{Thaler:2010tr}, and template
overlap \cite{Almeida:2010pa}.   The jet observable defined in the next section draws from the complimentary information  offered by color flow.  In a QCD event, radiation is controlled by the kinematics of the hard  partons as well as by how color indices are contracted together (color flow). Partons whose color indices are contracted together are color-connected, with a color string stretching between the two color sources.  For example, the two quarks in the hadronic  decay of a color singlet like the Higgs form a color dipole whose radiation pattern is contained primarily within a pair of cones around the two quarks, with a tendency for more radiation to occur in the region between the two quarks \cite{Bassetto:1984ik}.   

Color flow arguments of this sort have motivated attempts to use
QCD radiation patterns for event discrimination, {\it e.g.}\;mini-jet vetoes in Higgs searches \cite{Barger:1994zq}.  More recently, the authors of \cite{Gallicchio:2010sw} introduced a jet observable dubbed pull, which is a $p_T$-weighted vector in rapidity-phi space that is constructed so as to point from a given jet to its color-connected partner(s).  
Although pull has been shown to offer some discrimination in particle searches \cite{Black:2010dq}, it does not seem well-suited to 
tagging boosted hadronic tops.  
The most straightforward way to incorporate pull into a top tagging algorithm is 
to measure the pull of two subjets that reconstruct the $W^\pm$ and check whether
each subjet's pull vector points towards the other subjet.  A problem with this approach is that the pull vectors
are sensitive to how the $W^\pm$ jet is broken down into two subjets.  For a lopsided distribution of the 
$W^\pm$ into two subjets, one of the subjets will consist of only a small handful of calorimeter cells, and as a consequence
its pull will be sensitive to statistical fluctuations and contamination.  Even for a $W^\pm$ broken
down into two subjets more symmetrically, the pull vectors can depend sensitively on the precise boundary
drawn between the two subjets, which itself is a noisy function of the particular jet clustering 
algorithm being used.  Anticipating the potential difficulties of incorporating pull into a top tagging algorithm,
we explore an alternative approach in which the entire radiation pattern of the $W^\pm$ is
considered simultaneously.  This simple idea leads us to jet dipolarity, which we now define.  

\section*{Dipolarity}

Consider a jet, $J$, with two subjets, $j_1$ and $j_2$, whose centers are located at pseudorapidities $\eta_1$ and $\eta_2$ and azimuthal angles $\phi_1$ and $\phi_2$, respectively.  For each calorimeter cell ($\eta_i$, $\phi_i$) with transverse momentum $p_{Ti}$ let $R_i$ be the (minimum) euclidean distance 
in the $\eta$-$\phi$ plane between ($\eta_i$, $\phi_i$) and the line segment that runs from ($\eta_1$, $\phi_1$) to ($\eta_2$, $\phi_2$).  Dipolarity is defined as the $p_T$-weighted sum
\begin{equation}
\mathcal{D} \equiv \frac{1}{R_{12}^2 }\sum_{i\in J} \frac{p_{Ti}}{p_{T_J}}R_i^2
\label{eqn:dipdefn}
\end{equation}
where $R_{12}^2\equiv(\eta_1-\eta_2)^2+(\phi_1-\phi_2)^2$.  
Dipolarity is an infrared and collinear (IRC) safe observable so long as the algorithm used to identify $J$, $j_1$ and $j_2$ is IRC safe.
Notice that dipolarity, which is essentially a two-subjet observable, requires the centers
of $j_1$ and $j_2$ as input, although it does not require that the constituents of $J$ be partitioned
between $j_1$ and $j_2$.   The centers of $j_1$ and $j_2$ can be determined by whatever procedure
is convenient for the particular application.  For example one could choose the centers of $j_1$ and $j_2$ 
so as to minimize the sum in \eqref{eqn:dipdefn}. 

Dipolarity will be small when
most of the radiation within the jet $J$ occurs in the region between the two subjets $j_1$ and $j_2$ and will
be large whenever a substantial amount of radiation is found elsewhere.  As a consequence of the weighting with
respect to $R_i^2$ in \eqref{eqn:dipdefn}, $\mathcal{D}$ receives large contributions from semisoft radiation 
away from the cores of $j_1$ and $j_2$.  It is this semisoft radiation away from the cores of $j_1$ and $j_2$ that is expected to
reflect the color configuration of $J$.  The weighting in \eqref{eqn:dipdefn} does not know about the exact radiation pattern of
a color singlet; nevertheless, we expect that color singlets that decay into two jets will have small $\mathcal{D}$, while radiation
emitted by colored objects will tend to yield larger values of $\mathcal{D}$.  

\begin{figure}[h] %  figure placement: here, top, bottom, or page
   \centering
  \includegraphics[width=3.25 in]{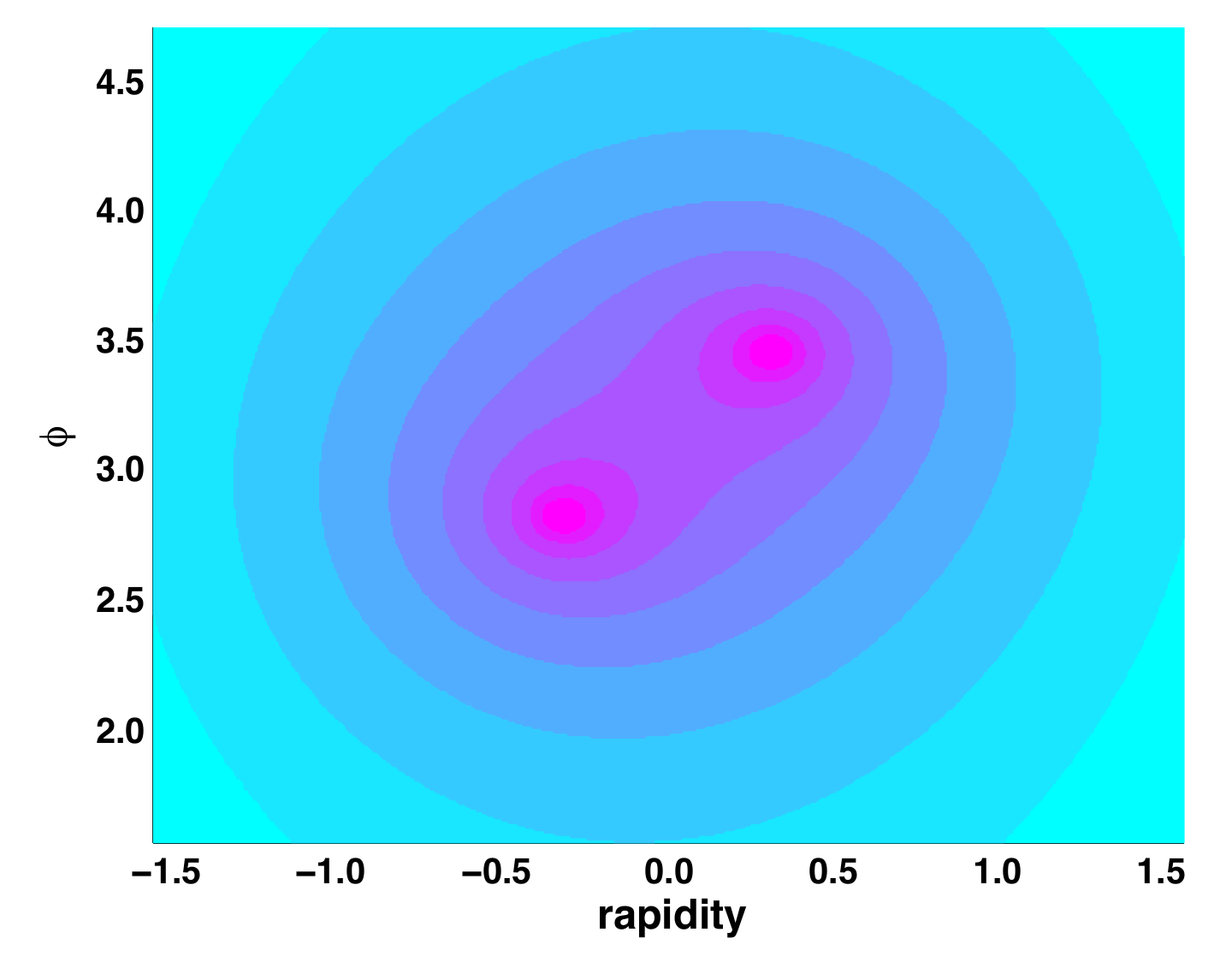}
  \includegraphics[width=3.25 in]{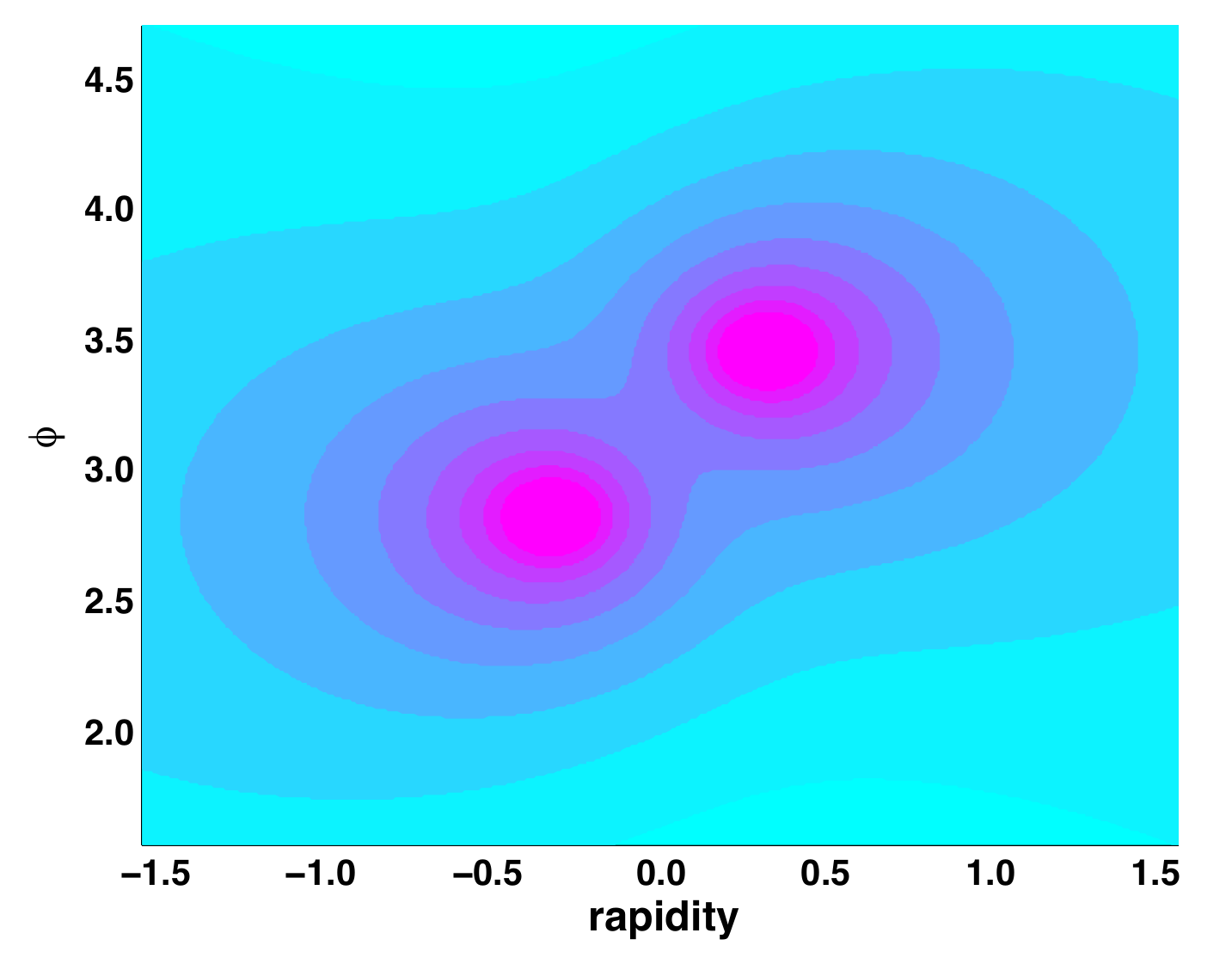}
   \caption{Top: Eikonal radiation pattern $dp_T/d\eta d\phi$ for a color singlet with $\Delta R$=0.9, typical for a $W^\pm$ originating from a 
   top with $p_T\!\sim$~300 GeV.  Bottom:  As above with the partons instead color-connected to the beam 
   (left/right-going parton connected to the left/right beam).  Contours differ by powers of $e$.  
   For the color singlet the radiation is mostly found in the region between the two subjets.  
   For the background-like color configuration, the radiation is pulled towards the beam.
   Note that an absolute comparison cannot be made
   between the figures, since the collinear singularities in \eqref{eqn:sigpat} and \eqref{eqn:bkgpat} are not regulated.}
   \label{fig:radpat}
\end{figure}

This expectation can be fleshed out more explicitly by considering the emission pattern of a third parton with energy $\omega$ from a pair of partons in a particular color configuration, see {\it e.g.}\;\cite{Bassetto:1984ik}.  
In the eikonal approximation ($\omega \to 0$) one finds the radiation function for a color singlet to be
\begin{equation}
\label{eqn:sigpat}
W_\text{s}(\eta,\phi) \sim \frac{dyd\phi}{\chi(\eta,\phi; \eta_1, \phi_1)\chi(\eta,\phi; \eta_2, \phi_2)}
\end{equation}
while for two partons color-connected to the beam we have instead
\begin{eqnarray}
\label{eqn:bkgpat}
W_{\text{ns}}(\eta,\phi) \sim && \frac{d\eta d\phi}{\chi(\eta,\phi; \eta_1, \phi_1)\chi(\eta,\phi; \eta_{\text{beam}})}+\\
 && \frac{d\eta d\phi}{\chi(\eta,\phi; \eta_2, \phi_2)\chi(\eta,\phi; \eta_{\text{beam}})} \nonumber
\end{eqnarray}
where
\begin{equation}
\chi(\eta,\phi; \eta_i, \phi_i) \equiv \cosh(\eta-\eta_i)-\cos(\phi-\phi_i)
\end{equation}
The resulting radiation patterns are depicted in FIG.\ref{fig:radpat}.  One sees explicitly that the color singlet has its radiation clustered in the region between the two partons, whereas for partons color-connected to the beam, a substantial amount of radiation  is emitted towards the beam.  
Using the expressions in \eqref{eqn:sigpat} and \eqref{eqn:bkgpat} to calculate $\mathcal{D}$ gives the prediction
$\mathcal{D}_{\text{ns}} \sim 2\,\mathcal{D}_{\text{s}}$; although this is approximately what is found from Monte Carlo calculations, 
expressions \eqref{eqn:sigpat} and \eqref{eqn:bkgpat} do not yield dipolarity distributions in quantitative agreement with the Monte Carlo.
Given the crudeness of the approximations that went into these expressions, this discrepancy is not surprising; a more accurate estimate
of $\mathcal{D}$ for various color configurations could be obtained by using antenna patterns as in \cite{Larkoski:2009ah}.

\begin{figure}[h] 
   \centering
  \includegraphics[width=3.25 in]{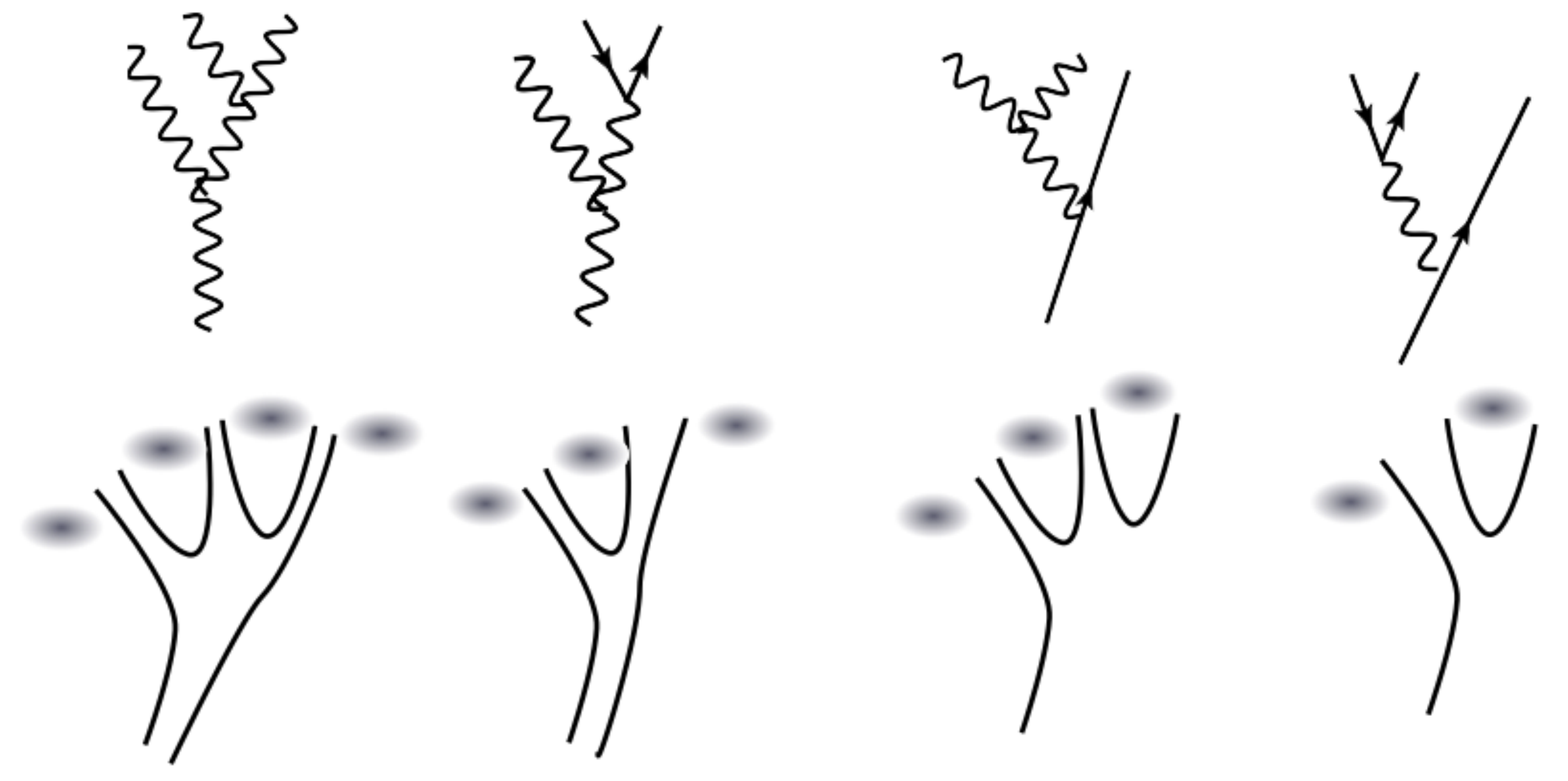}
   \caption{Schematic for a collection of QCD jets whose kinematics fake the top.  The upper figures show various possibilities for 
   quarks and gluons that undergo
   two branchings.  The bottom figures show the corresponding large $N_c$ color diagrams, with dipole radiation patterns superimposed across
   color dipoles.  Only the rightmost color configuration, which is suppressed by factors of $C_A/C_F$ with respect to the
   others, matches the radiation pattern of an actual top.}
   \label{fig:jetrad}
\end{figure}

Dipolarity can be used within the context of top tagging to reduce QCD backgrounds.  
Consider a collection of fat QCD jets originating from 
parton branchings with identical kinematics but different color configurations as illustrated in FIG.~\ref{fig:jetrad}.  
If one of the QCD jets fakes the kinematics of a top quark decay, then each of the different color configurations
fakes the kinematics equally well.  The dipolarities of the subjets, however, will be broadly distributed in accord with their
different color configurations.  For instance, gluon jets are known to give the largest fake rates for top jets as a consequence 
of their larger Casimirs which more often result in wide angle branchings with significant mass drops.  FIG.~\ref{fig:jetrad} illustrates
how gluon jets, with their distinct color configurations, radiate differently from top jets.
All of this suggests that the dipolarity of the $W^\pm$ in a hadronic top decay is well-suited 
as a discriminant in top tagging algorithms.

\section*{{\tt HEPTopTagger}}

To test whether dipolarity makes an effective discriminant,  cuts on dipolarity are incorporated into the {\tt HEPTopTagger }\cite{Plehn:2009rk,Plehn:2010st}, which
is designed to work effectively at intermediate boost, with $200 \GeV\!\lesssim\!p_T\!\lesssim\!800 \GeV$.  
The high efficiency  of the {\tt HEPTopTagger} at these $p_T$ makes it a good candidate for such a modification because dipolarity cuts are expected to be most effective at intermediate
$p_T$.  This is because at lower $p_T$ contamination from pile-up and the underlying event
becomes more of a concern as the top jets become fatter and fatter, while at higher $p_T$ the finite resolution
of the detector makes it difficult to get an accurate handle on radiation patterns.  Furthermore, the multibody filtering implemented
by the {\tt HEPTopTagger} results in accurate reconstruction of the $W^\pm$.  
The {\tt HEPTopTagger} algorithm is defined as follows:
\begin{enumerate}
\item Using the Cambridge/Aachen algorithm cluster the event into fat $R=1.5$ jets.
\item Break each fat jet $j$ into hard subjets using the following mass-drop criterion: undo the last stage of clustering to yield
two subjets $j_1$ and $j_2$ (with $m_{j_1}>m_{j_2}$), keeping both $j_1$ and $j_2$ if $m_{j_1}<0.8 m_{j}$ and otherwise dropping
$j_2$; repeat this procedure recursively, stopping when the $m_{j_i}$ drop below 30 GeV. 
\item Consider in turn all possible triplets of hard subjets.  First, filter\footnote{That is, recluster the constituents
of the triplet using the Cambridge/Aachen algorithm with jet radius $R_{\text{filter}}$. Filtering was introduced in \cite{Butterworth:2008sd}.} 
each triplet with a resolution $R_{\text{filter}} = \text{min}(0.3,\Delta R_{ij}/2)$.
Next, using the five hardest constituent subjets of the filtered triplet calculate the jet mass $m_{\text{filt}}$.  Finally, choose the triplet whose
$m_{\text{filt}}$ lies closest to $m_t$.  
\item Recluster the five filtered constituents chosen in step 3 into exactly three subjets $j_1$, $j_2$, and $j_3$ ordered
in descending $p_T$.  Accept the fat jet as a top candidate if it passes any of the following three pairs of mass cuts:
\belowdisplayskip=0pt
\abovedisplayskip=6pt
\begin{eqnarray*}
&i)& l_\text{cut}  \le \text{arctan } m_{13}/m_{12} \le 1.3 \text{ with } l_\text{cut} = 0.2	 \\
&i')&R_{\text{min}} \le \frac{m_{23}}{m_{123}} \le R_{\text{max}} \\
&ii)&R_{\text{min}}^2\left(1+\frac{m_{13}^2}{m_{12}^2}\right)\le1-\frac{m_{23}^2}{m_{123}^2} \le 
R_{\text{max}}^2\left(1+\frac{m_{13}^2}{m_{12}^2}\right) \\
&ii')&\frac{m_{23}}{m_{123}}\ge b_\text{cut} \text{ with } b_\text{cut} = 0.35\\
&iii)&R_{\text{min}}^2\left(1+\frac{m_{12}^2}{m_{13}^2}\right)\le1-\frac{m_{23}^2}{m_{123}^2} \le 
R_{\text{max}}^2\left(1+\frac{m_{12}^2}{m_{13}^2}\right) \\
&iii')&\frac{m_{23}}{m_{123}}\ge b_\text{cut} \text{ with } b_\text{cut} = 0.35 \\
\end{eqnarray*}
Here $R_{\text{min}} = 85\%\times m_W/m_t$ and $R_{\text{max}} = 115\%\times m_W/m_t$.
\item Finally, require that the total $p_T$ of the three subjets defined in step 4 be greater than 200 GeV.\footnote{The 
{\tt HEPTopTagger} does not make use of $b$-tagging, which is a natural extension to the algorithm that
can result in significant improvements in background rejection.
Since dipolarity cuts are orthogonal to b-tagging, we do not explore the use of b-tagging in this paper.}
\end{enumerate}

Dipolarity cuts are introduced into the {\tt HEPTopTagger} by modifying step 4 above.  For a top candidate
that has passed one of the three pairs of mass cuts we calculate the dipolarity of the $W^\pm$ as identified by the
mass cut: {\it e.g.}\;for a top candidate that satisfies ii) and ii') the $W^\pm$ is identified as $j_1+j_2$.  If more than 
one of the pairs of mass conditions is satisfied in step 4, we choose the smaller dipolarity.  We find that
this procedure performs better than calculating the dipolarity of the pair of subjets that reconstructs $m_W$ most accurately.  

In addition to introducing dipolarity cuts, we also make cuts on the filtered mass of the reconstructed top, $m_\text{filt}$, which
is not done in the original {\tt HEPTopTagger}, where the cuts have been chosen so as to avoid any explicit mass scales.  We introduce
cuts on $m_\text{filt}$ for two reasons.  The first is to improve background rejection.  The second and main reason is that we are interested 
in determining whether dipolarity cuts are essentially orthogonal to cuts on kinematic observables.  To do this we must ensure 
that the {\tt HEPTopTagger} is using a full compliment of kinematic cuts, including cuts on $m_\text{filt}$.  In a particular application,
cuts on $m_\text{filt}$ 
may be undesirable.  In that case, the inclusion of dipolarity cuts would result in a larger improvement of background rejection.

\begin{figure}[tb] %  figure placement: here, top, bottom, or page
   \centering
  \includegraphics[width=3.00 in]{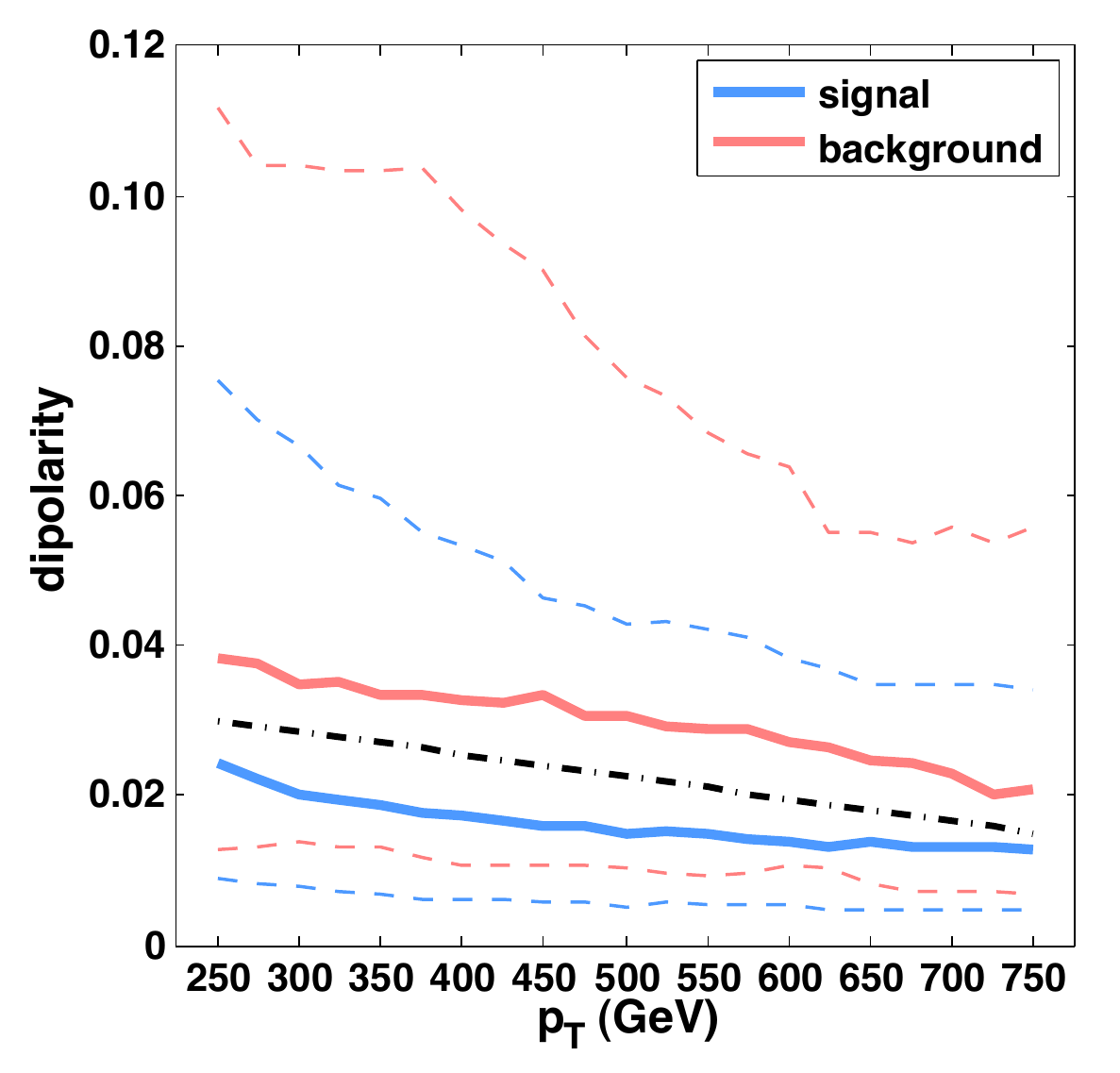}
   \caption{Dipolarity distributions for $W^\pm$s reconstructed by the {\tt HEPTopTagger} and passing default mass cuts with
   $m_{\text{filt}} \in$ [150 GeV, 210 GeV].  Thick solid lines indicate central values, whereas thin dashed lines 
   correspond to values at 10\% and 90\%.  Here and throughout the $p_T$ is that of the fat $R=1.5$ jet.
    For all $p_T$ the central value of the dipolarity for the background is
   $\OO(50\%-100\%)$ larger than for the signal.  This figure uses the {\tt HERWIG}  event samples; the {\tt PYTHIA} event samples yield
   similar distributions.  The dot-dash line roughly indicates where dipolarity cuts are made at the  S=20\% working point.}
   \label{fig:dipdistros}
\end{figure}

Note that the $j_i$ selected in step 4 contain only the hard substructure of the fat jet.  Some amount of soft
radiation has been thrown out by filtering and the mass drop criterion.  To effectively gauge whether
the radiation pattern of the reconstructed $W^\pm$ is consistent with the expected dipole radiation pattern, it
is important to include some of the discarded soft radiation.  We find
that the criterion used to select the radiation included in calculating the dipolarity of the $W^\pm$ has significant
impact on the ultimate utility of dipolarity as a discriminant.  In particular different criteria will lead
to dipolarity distributions that are more or less correlated with the kinematic observables considered by
the {\tt HEPTopTagger}.  Applying dipolarity in another context would likely require this criterion to be carefully
reworked so as to maximize performance. In the present case we find that the following criterion, which aims to capture as much of the radiation
emitted by the $W^\pm$ as possible, while minimizing possible contamination, to be most effective.  In addition to the hard radiation
from the two $W^\pm$ subjets, we include all soft radiation contained within the pair of cones centered around the 
two hard $W^\pm$ subjets, fixing the radius of the cones to be $\Delta R/\sqrt2$, where $\Delta R$ is defined
between the two hard $W^\pm$ subjets.  Furthermore we exclude any radiation contained within the smallest cone
that encloses the hard b subjet.  Note that angular ordering implies that the majority of the radiation emitted
by the $W^\pm$ is within the pair of cones of radius $\Delta R$.  We choose our cones to be somewhat smaller to minimize
contamination from underlying event/pile-up as well as the $b$ subjet.  See FIG.~\ref{fig:legoplot} to see this selection criterion 
at work on a sample top jet.  The orange cells in the figure correspond to radiation that has been discarded by mass-drop filtering
but that is included in calculating the dipolarity of the candidate $W^\pm$.

\vspace{10pt}
\section*{Testing Dipolarity}

In order to make a meaningful comparison between the performance of the {\tt HEPTopTagger} with and without dipolarity cuts, it is not enough to leave the kinematic cuts employed by the {\tt HEPTopTagger} fixed at their default values.  Instead it is important to optimize the cuts to yield 
the largest background rejection at each given signal efficiency.  
This optimization is performed by a custom Monte Carlo code that finely samples the space of cuts.  
Specifically, a scan is performed at discrete values of $R_{\text{min}}$ 
\begin{eqnarray*}
70\%\;m_W/m_t \le R_{\text{min}}\le   98\%\; m_W/m_t
 \end{eqnarray*}  
 with step sizes of $1\% \;m_W/m_t$.
 For each value of $R_{\text{min}}$, $R_{\text{max}}$ is chosen to be $R_{\text{max}} = 2.0\times m_W/m_t - R_{\text{min}}$.  
 In addition we simultaneously optimize over cuts $m_{t\text{ min}} \le m_\text{filt} \le m_{t\text{ max}}$,  $\mathcal{D}\le\mathcal{D}_{\text{max}}$, and, in step 4 of the {\tt HEPTopTagger}, $l_\text{cut} \le \text{arctan } m_{13}/m_{12}$ and   $b_\text{cut} \le m_{23}/m_{123}$.  
The remaining cuts in step 4 of the {\tt HEPTopTagger} are left at their default values, since these are less important for background rejection.  
Additionally,   the two parameters that define the mass drop criterion remain fixed at their default values.  The hard substructure cutoff of 30 GeV is sensitive to detector effects, which can only be crudely mocked-up without a full detector simulation.  See TABLE 
\ref{tab:operating point} for a sample set of cuts.  

We use three different event samples for evaluating the performance of the modified {\tt HEPTopTagger}.  
These event samples (with center of mass energy of 7 TeV) belong to a set of benchmark event samples that have been made publicly available by participants of BOOST 2010 \cite{BOOSTEvents}.  
The first event sample is generated by {\tt HERWIG}~6.510 \cite{herwig} with the underlying event simulated by {\tt JIMMY} \cite{jimmy}, which has been configured with a tune used by ATLAS.  
The second is generated by {\tt PYTHIA}~6.4 \cite{pythia} with $Q^2$-ordering and the `DW' tune for the underlying event.  The
third is generated by {\tt PYTHIA}~6.4 \cite{pythia} with $p_T$-ordering and the `Perugia' tune for the underlying event.  See 
\cite{Abdesselam:2010pt} for
more details.  For signal jets we use the hardest jet in each event of a Standard Model hadronic $t\bar t$ sample, excluding
jets with $|\eta| > 2.5$.  For background jets we use the hardest jet in each event of a Standard Model dijet sample, again excluding jets with $|\eta| > 2.5$.  
For jet clustering we use the Cambridge-Aachen (CA) algorithm \cite{Dokshitzer:1997in,Wobisch:1998wt} as implemented by  {\tt
FastJet~2.4.2} \cite{Cacciari:2005hq}.  
In order to simulate the finite resolution of the ATLAS or CMS calorimeters, particles in each event are clustered into $0.1 \times 0.1$ cells
in $\eta\!-\!\phi$ space and then combined into massless four-vector pseudoparticles that are fed into {\tt FastJet}.  We have also checked that
imposing a low energy cutoff of 1 GeV on each cell results in only a mild degradation of background rejection.

\begin{figure}[ht] %  figure placement: here, top, bottom, or page
   \centering
  \includegraphics[width=3.25 in]{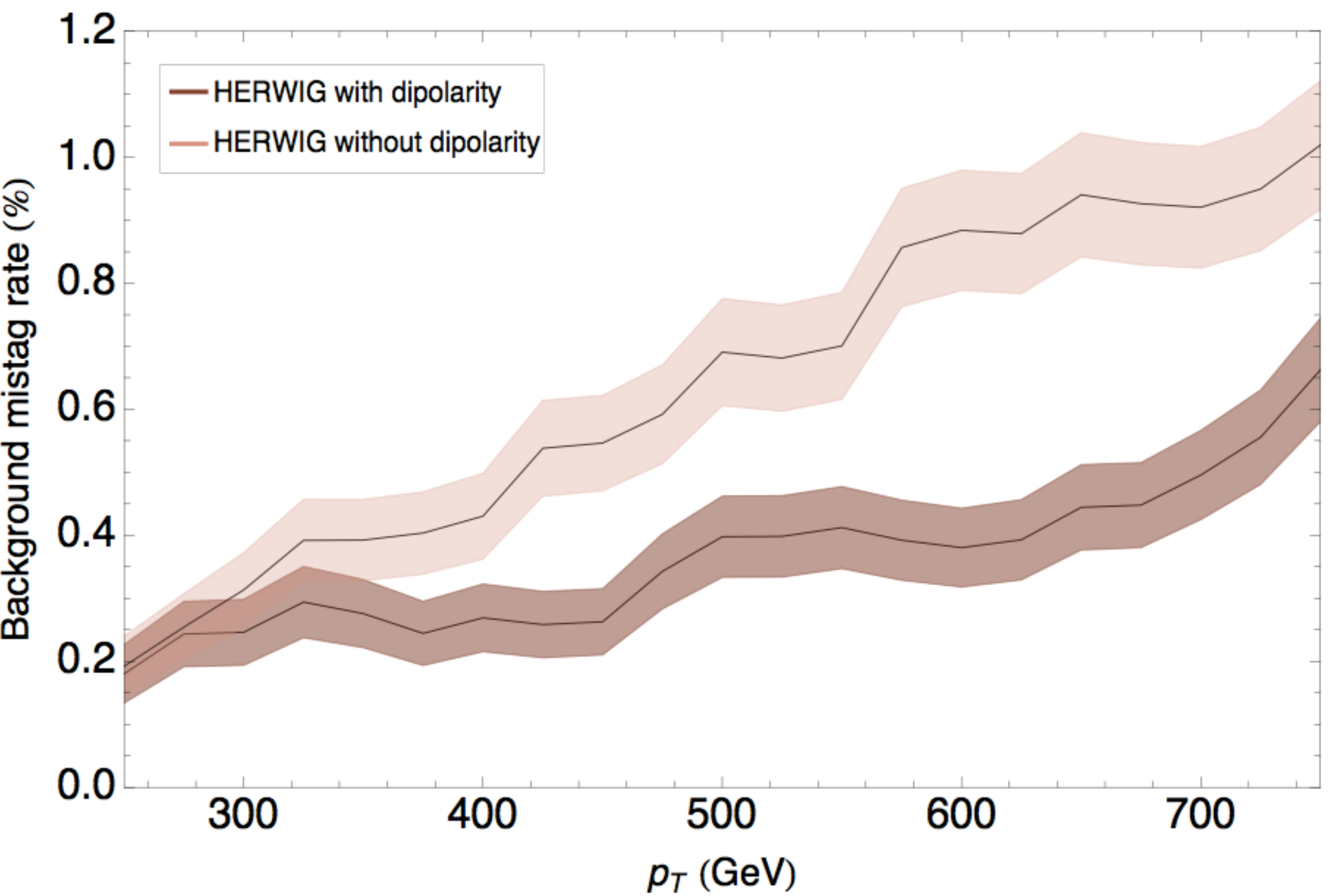}
    \includegraphics[width=3.25 in]{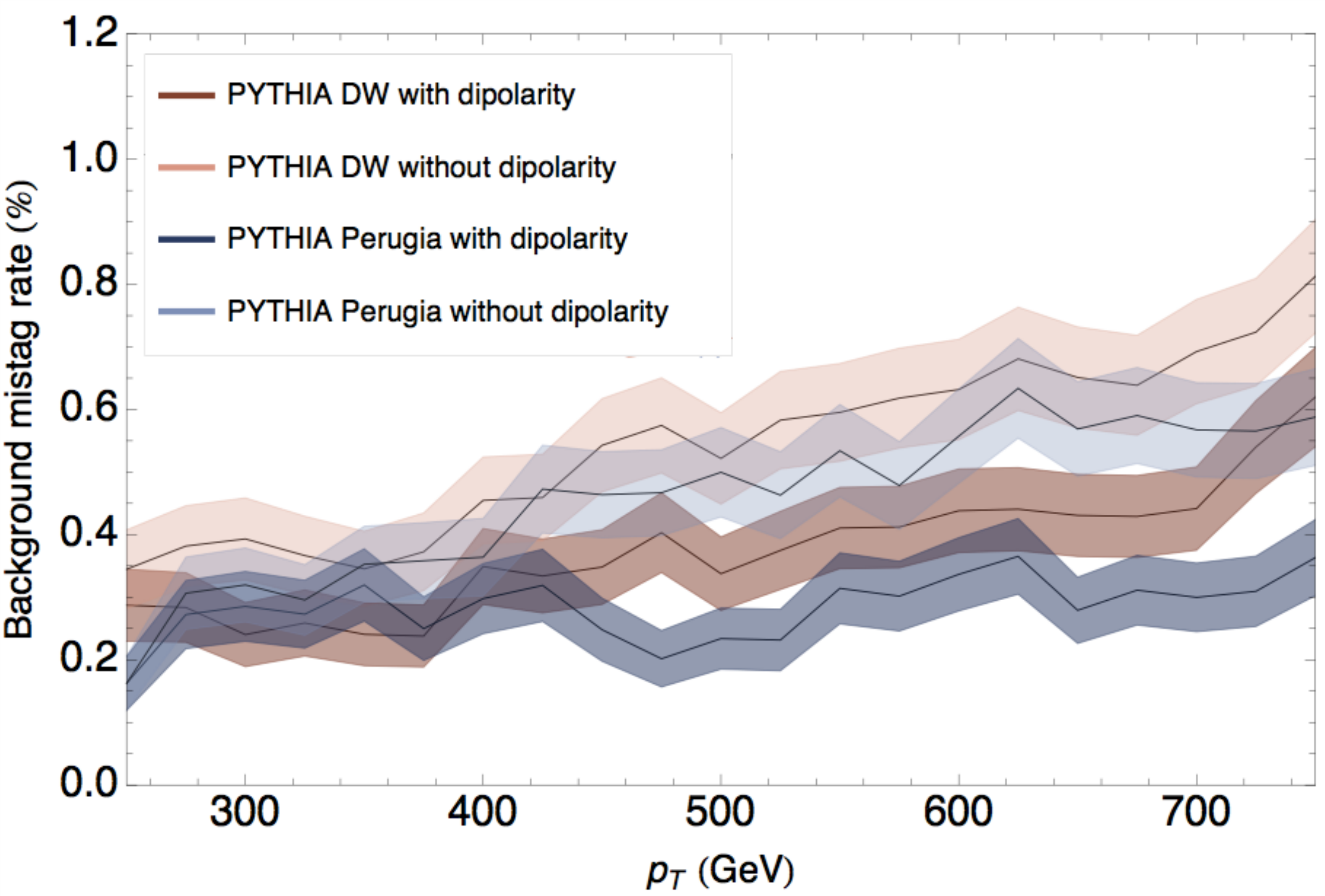}
   \caption{Background mistag rates with and without dipolarity cuts at a fixed signal efficiency
   of 20\% for {\tt HERWIG} (top) and {\tt PYTHIA} (bottom).  The mistag rate at a given $p_{T0}$ is calculated from a $p_T$ window
   of 100 GeV centered at $p_{T0}$. Note that, as a consequence, each point is not statistically independent.  Error bands are statistical.}
   \label{fig:finept}
\end{figure}
  \begin{figure}[ht] %  figure placement: here, top, bottom, or page
   \centering
  \includegraphics[width=3.5 in]{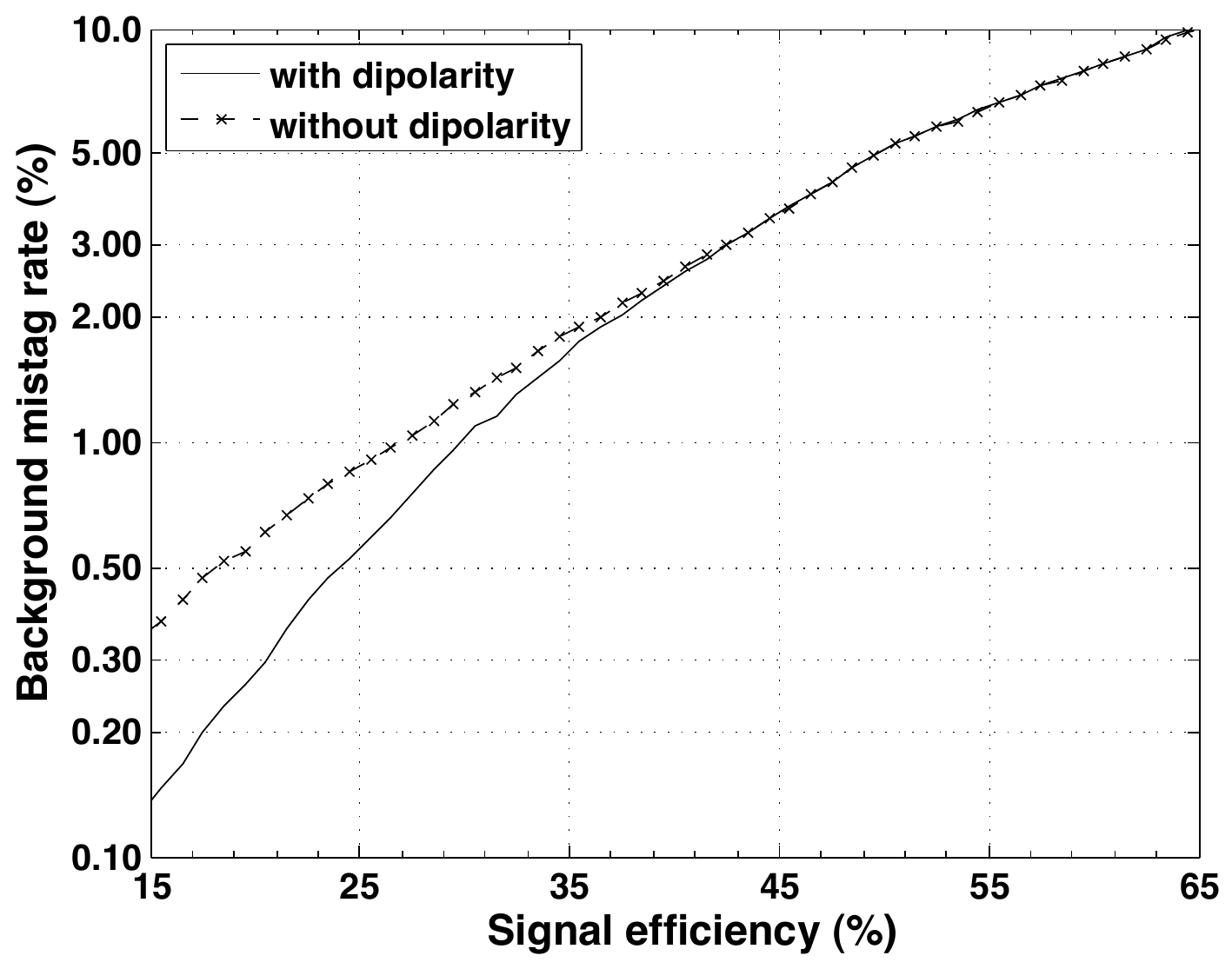}
   \caption{Signal efficiency vs.\;background mistag rate for the {\tt HEPTopTagger} with $p_T \in [400 \GeV, 500 \GeV]$ and {\tt HERWIG} event 
   samples.  At lower signal
   efficiencies the inclusion of dipolarity cuts leads to a sizeable improvement in background rejection.  Statistical error bars, which are a relative
   25\% at the lowest mistag rate, are not shown.  }
   \label{fig:herwigbscurve}
\end{figure}
 
%\section*{Results} 
 
 \begin{table*}[th]
\begin{center}
\begin{tabular}{|l||c|c|c|c|c|c||c|c|c|}
\hline
%Optimized 
\multicolumn{1}{|c|}{Operating Point}
&$l_{\text{cut}}$& $b_{\text{cut}}$& $r_{\text{min}}$& $\mathcal{D}_{\text{max}}$& $m_{t\text{ min}}$& $m_{t\text{ max}}$&\multicolumn{3}{|c|}{$B$ (\%)}
\\
\hline
\hline
Low $p_T$ without $\mathcal{D}$ cut& 0.45 & 0.41 &0.92& - & 159 GeV & 195 GeV&0.47 &0.48 &0.41\\
\hline
Low $p_T$ with $\mathcal{D}$ cut& 0.37 & 0.39 &0.80& 0.021& 154 GeV & 199 GeV&0.41& 0.38 &0.30\\
\hline
\hline
High $p_T$ without $\mathcal{D}$ cut& 0.47 & 0.40 &0.93& - & 158 GeV & 199 GeV&0.92 &0.79 &0.59\\
\hline
High $p_T$ with $\mathcal{D}$ cut& 0.36 & 0.38 &0.88& 0.023 & 154 GeV & 196 GeV&0.58&0.58 &0.39\\
\hline
\end{tabular}
\caption{
Sample optimized operating points at $S=20\%$ based on an equal admixture of all three event samples for maximum statistics.  
The resulting background mistag rates (B) are shown for each of the three event samples with {\tt HERWIG}, {\tt PYTHIA} `DW', and 
{\tt PYTHIA} `Perugia' arranged from left to right. Including dipolarity cuts loosens mass cuts while improving background rejection.  
The low $p_T$ samples have $200 \GeV \le p_{T} \le 500 \GeV$ while the high $p_T$ samples have $400 \GeV \le p_{T} \le 800 \GeV$.
\label{tab:operating point}
}
\end{center}
\end{table*}
 
FIG.~\ref{fig:dipdistros} shows the dipolarity distributions that result from the modified {\tt HEPTopTagger} for a default set of mass
cuts.  As expected the dipolarity of the reconstructed $W^\pm$ in top jets is smaller than for the QCD background.
FIG.~\ref{fig:finept} shows the improvement in performance that results from including dipolarity cuts into the {\tt HEPTopTagger} as
a function of $p_T$ at a fixed signal efficiency of $S=20\%$.
Finally, FIG.~\ref{fig:herwigbscurve} shows how, for $p_T \in [400 \GeV, 500 \GeV]$, dipolarity cuts improve background rejection at
signal efficiencies $S\lsim 35\%$.

At a fixed signal efficiency of $S=20\%$, dipolarity cuts lead to a sizable decrease in the mistag rate for $350 \GeV\!\lesssim\!p_T\!\lesssim\!800 \GeV$ with the largest decrease at intermediate $p_T$.  
For the {\tt HERWIG} event samples the mistag rate decreases by as much as $\sim 50\%$,
whereas for the {\tt PYTHIA} samples the mistag rate decreases by a more modest amount, $ \sim \!30\%$.  
Differences in the underlying event (UE) model do not explain this disagreement; for instance, 
repeating the analysis with the UE turned off results in
background mistag rates that are only somewhat smaller, with both {\tt HERWIG} and {\tt PYTHIA} dropping by similar amounts.  
This suggests that the difference between the {\tt HERWIG} and  {\tt PYTHIA} mistag rates most likely arises from the parton shower.

\section*{Discussion}

This article has introduced a new jet substructure observable that is useful in discriminating among different color configurations in jets that have large mass drops.  This discrimination is of interest, since such jets often arise from decays of boosted heavy particles.  Incorporating this
discriminant into a top tagging algorithm results in QCD background mistag rates that are
lower by as much as 50\%; the exact mistag rate, however, shows considerable sensitivity to the details of the parton shower.
Specifically {\tt HERWIG} event samples result in a larger improvement in background rejection than is found for {\tt PYTHIA}.  
We suspect that {\tt HERWIG}, which uses angular ordering, does a better job of simulating the effects of color coherence than
{\tt PYTHIA}, which uses $Q^2$ or $p_T$-ordering in combination with angular vetoes.  This could explain why the dipolarity of the $W^\pm$
is a more discriminating observable in the case of the {\tt HERWIG} event samples.   
With an understanding of the origin of this difference, comparisons to measurements at the LHC could help improve the description of QCD radiation. 
It would be interesting to understand how this difference arises from the details of the parton shower;
doing so, however, lies outside the scope of this paper.

Validating how well color flow effects as modeled by Monte Carlo event generators match what is observed in collider
experiments is only beginning to be studied actively.
Understanding color flow in detail is a difficult problem; for example, QCD predictions for radiation patterns can be affected by non-global logarithms, see {\it e.g.}\;\cite{Dasgupta}.  Therefore validating theoretical predictions against data will be critical in reducing the theoretical uncertainty associated with how dipolarity and other color flow observables are modeled by Monte Carlo calculations.  
A few color coherence studies performed at the Tevatron showed spatial correlations between the third and second hardest jets in $p\bar{p}$ collisions, and {\tt HERWIG} was shown to provide a better description of the data than {\tt PYTHIA} \cite{Varelas:1998aw}.
More recently, the color of the $W^\pm$ in $t\bar{t}$ events was studied, and agreement between theory predictions for jet pull and data was shown \cite{Abazov:2011vh}.

Jet dipolarity should be useful in a broader set of applications to both Standard Model and beyond the Standard Model physics.
%specifically to search channels in which a heavy particle decays into lighter color singlets (or is itself a color singlet) that ultimately decay into jets.
Possible directions for future research include: 
(i) applications of dipolarity to a collider search for heavy color singlet resonances that decay to $t\bar t$;
(ii) applications to standard model $W^\pm/Z^0$ physics; 
(iii) applications to heavy color singlet resonances that decay to $W^+W^-$ or $Z^0Z^0$; 
(iv) applications to cascade decays of supersymmetric particles;
 (v) inclusion of dipolarity into other top-tagging algorithms;
 (vi) applications to the decay of new particles into novel color configurations such as in the decay of the LSP in supersymmetric models 
 with baryonic R-parity violation; and
 (vii) modifying $\mathcal{D}$ to more closely correspond to the exact radiation pattern of a color singlet.
 Each of these directions could make an interesting laboratory for further development of jet substructure techniques.

\vspace{10pt}

\section*{Acknowledgements}
We would like to thank Maria Baryakhtar for collaboration at early stages of this work.  
We would also like to thank Tilman Plehn, Gavin Salam and Michael Spannowsky for providing us with their implementation of the {\tt HEPTopTagger}.   JW would like to thank Tilman Plehn, Gavin Salam, and David E. Kaplan for useful conversations during the course of this work.  MJ would like to thank Jason Gallicchio for interesting conversations about color pull.
MJ, AH and JGW are supported by the US DOE under contract number DE-AC02-76SF00515.  
MJ, AH and JGW receive partial support from the Stanford Institute for Theoretical Physics.  
JGW is partially supported by the US DOE's Outstanding Junior Investigator Award and the Sloan Foundation.

%%%%%%%%%%%%%%%%%%%%%%%%%%%%%

\end{document}